# The emitted energy in RS

I.V. Dovgan*

Department of Physics, Moscow State Pedagogical University, Moscow 119992, Russia.

*dovganirv@gmail.com

**Abstract:** The emitted energy of electrons is calculated in relativistic strophotron. It is shown that it is superposition of amplification at harmonics of main resonance frequency of the system.

## 1. Introduction

Free electron lasers (FEL) are powerful coherent sources of radiation [1,2]. The main element of FEL is undulator in which electrons are oscillating in transverse direction. There are many other types of FEL [3-39] and references therein. One of them is relativistic strophotron where electrons move in parabolic potential through.

In the present article the gain is calculated in relativistic strophotron.

## 2. The emitted energy

We start from the following equations of motion of the electron in the electrostatic field with the scalar potential $\Phi(x)=\Phi_0(x/d)^2$ and in the electromagnetic wave propagating along the *z* axis with frequency $\omega$ and electric field strength amplitude $E_0$:

$$\begin{aligned} \frac{dp_z}{dt} &= ev_x E_0 \cos\omega(t-z), \\ \frac{dp_x}{dt} &= eE_0\left(1-v_z\right)\cos\omega(t-z)-\frac{2e\,\Phi_0}{d^2}x. \end{aligned} \tag{1}$$

The total (kinetic + potential) energy of the electron is

$$\varepsilon_{tot}=\left(m^2+p_x^2+p_y^2\right)^{1/2}+e\,\Phi_0\frac{x^2}{d^2} \tag{2}$$

and the rate of its change is equal to [41]

$$\frac{d\varepsilon_{tot}}{dt} = eE_0 v_x \cos\omega(t - z). \tag{3}$$

In this paper, we will only investigate the linear gain to find which is sufficient to obtain the first-order corrections $x^{(1)}$ and $z^{(1)}(t)$ to $x^{(0)}(t)$ and $z^{(0)}(t)$ given by (1) and (3) of [42]. These first-order corrections obey the equations

$$\begin{cases} \dfrac{dp_x^{(1)}}{dt} = -\varepsilon_0 \Omega^2 x^{(1)} + eE_0 \cos\omega\left(t - z^{(0)}\right), \\ \dfrac{dp_z^{(1)}}{dt} = eE_0 v_x^{(0)} \cos\omega\left(t - z^{(0)}\right). \end{cases} \tag{4}$$

The linear (field-independent) gain is determined by the second-order $\left(\propto E_0^{\ 2}\right)$ part of $\dfrac{d\varepsilon_{tot}}{dt}$ (3) which can be expressed in terms of $x^{(1)}$, $z^{(1)}$

$$\frac{d\varepsilon_{tot}}{dt} = eE_0 \left\{ \dot{x}^{(1)} \cos\omega(t - z^{(0)}) + \omega v_x^{(0)} z^{(1)} \sin\omega(t - z) \right\}. \tag{5}$$

The energy emitted by the electron during the time T is given by

$$\Delta\varepsilon = -\int_0^T \frac{d\varepsilon_{tot}}{dt} dt. \tag{6}$$

On the right-hand side of the first equation (4), the cosine of the periodical function in $z^{(0)}(t)$ (3) of [42] can be expanded in the Fourier series to give the Bessel functions with the argument $Z$ (8) of [42]. Then the first equation (4) is easily integrated to give $p_z^{(1)}$.

To integrate the second equation (2), we first use the relations

$$\begin{aligned} p_x &= v_x \frac{\sqrt{p_z^2 + m^2}}{\sqrt{1 - v_x^2}}, \\ \frac{dp_x}{dt} &= \frac{dv_x}{dt} \frac{\sqrt{p_z^2 + m^2}}{\left(1 - v_x^2\right)^{3/2}} + v_x \frac{p_z dp_z / dt}{\sqrt{1 - v_x^2}\sqrt{p_z^2 + m^2}}. \end{aligned} \tag{7}$$

Finding from the second equation (7) $\dfrac{dp_x^{(1)}}{dt}$ and then using the condition $v_x << 1$, we reduce the second equation (4) to the form

$$\ddot{x}^{(1)} + \Omega^2 x^{(1)} = \frac{eE_0}{\varepsilon_0}\left(1 - v_z^{(0)}\right)\cos\omega\left(t - z^{(0)}\right) - \frac{1}{\varepsilon_0}\frac{d}{dt}\left(v_x^{(0)} p_z^{(1)}\right). \tag{8}$$

Using $p_z^{(1)}$ found from the first equation (4), we can integrate (8) to find $x^{(1)}$. Now the correction $v_z^{(1)}$ to $\dot{z}^{(0)}(t)$ (3) of [42] can be found from the relation

$$v_z = \frac{p_z}{\varepsilon_z}\left(1 - \frac{1}{2} v_x^2\right)$$

which, in the first order, takes the form

$$v_z^{(1)} \approx \frac{p_z^{(1)}}{\varepsilon_0 \gamma^2} - v_x^{(0)} v_x^{(1)}. \tag{9}$$

In principle, all of these calculations are simple, although rather cumbersome. They will be described in detail elsewhere. Here we will present only the found result:

$$\Delta\varepsilon = \frac{e^2 E_0^2 T^3 \omega}{64\varepsilon_0}\left(\alpha^2 + x_0^2\Omega^2\right)\left(\frac{1}{\gamma^2} + \frac{\alpha^2 + x_0^2\Omega^2}{4}\right) \times \sum_s \frac{d}{du_s}\frac{\sin^2 u_s}{u_s^2}\left(J_s(Z) - J_{s+1}(Z)\right)^2. \tag{10}$$

Here $u_s = \frac{T}{4\gamma^2}\left[\omega\left(1 + \frac{\gamma^2}{2}\left(\alpha^2 + x_0^2\Omega^2\right)\right) - 2\gamma^2\Omega(2s+1)\right]$ and $Z = \frac{\omega}{8\Omega}\left(\alpha^2 + x_0^2\Omega^2\right)$.

The same result in the same manner can be derived from the classical equations of motion in the case of the strophotron with a magnetic "trough." The result (10) was also derived using a different approach in the framework of a quantum-mechanical description of electron motion [8].

To compare $\Delta\varepsilon$ (10) in the strophotron to the corresponding result for FEL with the plane undulator [2], we will rewrite (10) in a different form using the effective "undulator parameter" $K = K_{str} = \gamma\,\Omega a$:

$$\Delta\varepsilon = \frac{e^2 E_0^2 T^3 \Omega}{32\varepsilon_0 \gamma^2} K^2 \frac{1 + \frac{1}{4}K^2}{1 + \frac{1}{2}K^2} \times \sum_s (2s+1)\frac{d}{du_s}\frac{\sin^2 u_s}{u_s^2}\left(J_s(Z) - J_{s+1}(Z)\right)^2. \tag{11}$$

In the case of the plane undulator, the results derived by Becker [2] and Colson [2] written in the same form are given by

$$\Delta\varepsilon = \frac{e^2 E_0^2 T^3 \Omega}{16\varepsilon_0 \gamma^2} K^2 \sum_s (2s+1)\frac{d}{du_s}\frac{\sin^2 u_s}{u_s^2}\left(J_s(Z) - J_{s+1}(Z)\right)^2. \tag{12}$$

Where

$$K = K_{und} = \frac{eB_0\lambda_0}{2\pi mc^2}.$$

A comparison of (11) and (12) shows that there is a difference determined by the factor

$$\frac{1}{2}\frac{1 + \frac{1}{4}K^2}{1 + \frac{1}{2}K^2} \equiv \xi$$

Numerically, this difference is not too important because $\frac{1}{4} \le \xi \le \frac{1}{2}$. But, on the other hand, this result shows that there is not a complete coincidence between the strophotron and the undulator, even at the single-electron level. This difference becomes much more pronounced when we consider the beam as a whole because of a difference in the definitions of the $K$ parameter in the undulator and in the strophotron. In the latter case, the parameter $K$ depends on $x_0$, whereas in the case of the undulator, the $K$ parameter is $x_0$ independent.

Each term of the sums (10), (11) describes a stimulated emission in the vicinity of the 2s + 1-st harmonic of the main resonance frequency $\omega_{res} = \frac{2\gamma^2\Omega}{1+\frac{\gamma^2}{2}\left(\alpha^2 + x_0^2\Omega^2\right)}$.

If $\alpha$ and $x_0$ are fixed, i.e., for a single electron, the lines of stimulated emission do not overlap because they are separated by the term $\omega_{res}$ exceeding their homogeneous width $\omega_{res} / \Omega T$ where $\Omega T >> 1$. The gain at higher harmonics is determined by the factor $\left(J_s(Z) - J_{s+1}(Z)\right)^2$. which is not small only if $\gamma^{?}\left(\alpha^2 + x_0^2\Omega^2\right) \equiv K_{str} >> 1$ . In this case, the resonance frequency $\omega_{res} = \frac{2\gamma^2\Omega}{1+\frac{\gamma^2}{2}\left(\alpha^2 + x_0^2\Omega^2\right)}$ differs essentially from the resonance frequency 27% found in [3]. This difference indicates once more that the results of [3] are correct only for $K < 1$, and hence they are applicable only for a description of amplification at the main frequency $\omega_{res}$.

## 3. Conclusion

The amplification spectrum in relativistic strophotron is shown to have a form of superposition of many contributions of harmonics of main resonance frequency of the system. The obtained results are compared with results of usual FEL on undulators. It is shown, that for single electron the difference of gains in strophotron and plane undulator is no sufficient.